\documentstyle[aps,twocolumn,epsfig]{revtex}
\begin{document}
\draft

\twocolumn[\hsize\textwidth\columnwidth\hsize\csname@twocolumnfalse%
\endcsname

\title{A non-extensive approach to the time evolution of Lyapunov
coefficients}
\author{Massimiliano Ignaccolo$^{1}$, Paolo Grigolini$^{1,2,3}$}
\address{$^{1}$Center for Nonlinear Science, University of North Texas,\\
P.O. Box 5368, Denton, Texas 76203 }
\address{$^{2}$Istituto di Biofisica CNR, Area della Ricerca di Pisa,\\
Via Alfieri 1,\\
San Cataldo 56010 Ghezzano-Pisa, Italy }
\address{Dipartimento di Fisica dell'Universit\'{a} di Pisa and INFM,\\
Piazza\\
Torricelli 2, 56127 Pisa, Italy }
\date{\today}
\maketitle

\begin{abstract}

We study sporadic randomness
by
means of a non-extensive form of Lyapunov coefficient. We recover from a
different perspective the same conclusion as that of an earlier work,
namely, that the ordinary Pesin theorem applies (P.Gaspard and X.-J.
Wang, Proc. Natl. Acad. Sci. USA {\bf85}, 4591 (1988) ).
However, our theoretical analysis allows us to organize the numerical
calculations so as to reveal the slow transition from a temporary
form of non-extensive thermodynamics, corresponding to the prediction
of a recent paper ( M. Buiatti, P. Grigolini, A. Montagnini,
Phys. Rev. Lett {\bf 82}, 3383 (1999)), to the ordinary extensive
thermodynamics. We show that the transition takes place with a slow decay
corresponding to the regression from a
non-equilibrium initial condition to equilibrium condition.

\end{abstract}

\bigskip
\mbox{}
]

\section{introduction}

The complex and erratic behavior of chaotic trajectories
is closely related to the so called sensitive dependence
on initial conditions. The degree of this sensitivity
is measured by the Lyapunov exponent \cite{hilborn} and this is the
reason why the Lyapunov exponent is so popular in the field of
deterministic chaos. The Lyapunov coefficient establishes, so to speak,
the rate of time increase of the function $\xi(t)$ defined by:
\begin{equation}
\xi(t)  \equiv  \lim_{\Delta x(0) \rightarrow  0}
\frac {\Delta x(t)}{\Delta x(0)}.
\label{delta}
\end{equation}

The meaning of this function is as follows. We consider a given trajectory
moving from the initial condition $x(0)$, and we
refer to it as \emph{trajectory of interest}. Then we study another
trajectory,
departing from an initial condition very close but distinct from that
of the trajectory of interest. We refer to this new trajectory as
\emph{auxiliary trajectory}. The quantity $\Delta x(t)$ denotes the
distance between the auxiliary trajectory and the trajectory of
interest at time $t$. Consequently, $\Delta x(0)$ denotes the
distance between the two distinct initial conditions.
The ordinary Lyapunov perspective rests on the assumption that:
\begin{equation}
\xi(t) = exp(\lambda t),
\label{conventional}
\end{equation}
where $\lambda$ denotes the conventional Lyapunov coefficient.

The adoption of the non-extensive
thermodynamics advocated by Tsallis \cite{brazil}, however, shows that
the exponential sensitivity behind the definition of the conventional
Lyapunov coefficient is a special case of a more general condition.
In fact, according to \cite{TPZ97},\cite{costa} and \cite{lyra},
the conventional form of Eq.({\ref{conventional}) has to be replaced by:
\begin{equation}
\xi(t) = [ 1 + (1-Q) \lambda_{Q}t]^{\frac{1}{1-Q}},
\label{unconventional}
\end{equation}
where $\lambda_{Q}$ denotes the generalized (non-extensive) Lyapunov
coefficient.

The index $Q$ is referred to as \emph{entropic index}\cite{brazil},
since the authors of \cite{TPZ97},\cite{costa} and \cite{lyra}
prove the structure of Eq. (\ref{unconventional}) to be a natural
consequence
of the non-extensive form of entropy proposed in 1988 by Tsallis
\cite{CONSTANTINO88}, which is expressed in fact as a function of $Q$ as follows:
\begin{equation}
H_{Q} = \frac{1 - \sum_{i} p_{i}^{Q}}{Q-1}.
\label{entropyt}
\end{equation}
We note that when the entropic index Q gets
the value $Q = 1$, the new structure of Eq. (\ref{unconventional}) becomes equivalent to the
ordinary structure of Eq. (\ref{conventional}) thereby showing that the
ordinary Lyapunov exponent must be identified with $\lambda_{1}$.

The purpose of this paper is to adapt the Lyapunov approach to this new perspective with a special attention to the case of an intermittent map of the same kind as that behind the dynamic approach to L\'{e}vy processes of ref. \cite{ANNA}. The authors of  \cite{ANNA} proved that the dynamics within the laminar region of the intermittent map adhere to the prescription of Eq. (\ref{unconventional}) with $Q>1$. This means that the function $\xi (t)$ of Eq. (\ref{unconventional}) diverges at a finite time, thereby implying an exit from the laminar into the chaotic region at much earlier times. Note that  the trajectory leaving the  laminar condition  enters  a
region characterized by Bernouilli randomness, and so by ordinary
Lyapunov coefficients. This means, in other words, that the resulting dynamics  are a balance between a regular motion described by Eq. (\ref{unconventional}) and a random motion corresponding to the prescription of Eq. (\ref{conventional}). Although this random motion is sporadic, it has the important effect of erasing the memory of the initial condition \cite{mauro}. We want to discuss the "thermodynamic" consequence of this sporadic randomness: This is the process of memory erasure necessary for the resulting  diffusion process to become equivalent to an ordinary L\'{e}vy process \cite{mauro}.

The outline of this paper is as follows. Section II is devoted to illustrating,  with the help of heuristic arguments, two distinct prescriptions to evaluate the Lyapunov coefficient in the non-extensive case.  In Section III
we supplement  the heuristic choice of the proper form of Lyapunov coefficent by using the generalized version of the Pesin theorem \cite{pesin}  established in an earlier work \cite{jin}. In Section IV we shall show that the assumption of the existence of a smooth invariant distribution implies the long-time dynamics of the intermittent map here under study, namely, the Maneville map \cite{manneville}, to be extensive. In Section V we study the transition from an off-equilibrium  condition to the natural invariant distribution: The corresponding time evolution of the Lyapunov coefficients is monitored by using a numerical approach. The physical significance of the results of this paper is illustrated in Section VI.

\section{Two distinct prescriptions to evaluate Lyapunov coefficients}

Let us now discuss two different approaches to the time evolution of  Lyapunov
coefficient, which turn out to be equivalent the one to the other only in the extensive case. For simplicity we shall limit our discussion to the one-dimensional case, and more specifically, to the case where  the phase space is given by  the interval $[0,1]$. Let us assume that
\begin{equation}
x_{n+1} = \Phi(x_{n}) .
\label{genericmap}
\end{equation}
We shall see that the two distinct definitions rest on the use of
only one auxiliary trajectory and of a conveniently
large number of them,
respectively. The first definition, therefore, seems to be closer to
the spirit of the function $\xi (t)$ of Eq. (\ref{delta}).

We discuss first the case of only one auxiliary trajectory. At any
time step we can define the corresponding conventional Lyapunov
coefficient as
\begin{equation}
\lambda_{n} \equiv ln \bigg| \, \frac{\Delta x(n+1)}{\Delta x(n)} \, \bigg|.
\label{ordinary}
\end{equation}
Thus we can define also the time dependent Lyapunov coefficient:
\begin{equation}
\Lambda(N,x_{0}) = \sum_{n=0}^{N-1}\lambda_{n}.
\label{SUM}
\end{equation}
We note that this Lyapunov coefficient can also be
written under the form:
\begin{equation}
\Lambda(N,x_{0}) = ln \prod_{n=0}^{N-1} \mid \Phi^{'}(x_{n})\mid ,
\label{product}
\end{equation}
where $\Phi^{'}(x)$ denotes the derivative of $\Phi (x)$ with respect to $x$.
We leave explicit the dependence of $\Lambda (N,x)$  on the initial condition
$x_{0}$ on purpose. In fact, in this paper we are interested in the physical effects caused by this form of memory, stemming from the laminar region of the phase space. As we shall see in Section IV, the phase space of the Manneville map is divided into two parts, a laminar region, responsible for these memory effects, and a chaotic region. A generic trajectory explores the chaotic region only from time to time, thereby making sporadic the action of randomness. This sporadic action produces the interesting effect of erasing the memory of the laminar region, and we plan to discuss the "thermodynamic" consequences of this process of memory erasure. In the simple case where the whole space phase is random and mixing, we expect, in agreement with the Pesin theorem \cite{schuster}, that the system reaches quickly the steady condition:
\begin{equation}
h_{KS} =  lim_{N \rightarrow \infty} \frac{\Lambda(N,x_{0})}{N}
= \frac {\int_{0}^{1} ln \mid \Phi^{'}(x) \mid \rho(x) dx}
{\int_{0}^{1}\rho(x) dx}.
\label{pesin}
\end{equation}
However, there are cases where it takes extremely long time for
this steady condition to settle. One of these cases is given by the
Manneville map \cite{manneville}. This map, which is the dynamical system
studied in this paper, is the prototype of sporadic dynamical
processes, and its algorithmic complexity has been discussed by
Gaspard and Wang \cite{gaspard}.

Let us now illustrate how to define the Lyapunov coefficient by means
of a conveniently large number of auxiliary trajectories. This is the procedure
adopted in the paper by Benettin and Galgani \cite{benettin}. We run
first of all the trajectories of interest moving from the initial
condition $x_{0}$. Then, we have recourse to a first auxiliary
trajectory, departing from a different initial condition, at a distance
$\Delta x(0)$. We move in time by one step. The  distance between the
two trajectories is now $\Delta x(1) > \Delta x(0)$. We artificially
reduce this new distance to the earlier distance, denoted by the
symbol $d$. Then we start a new auxiliary trajectory, and so on. It
is evident that this prescription yields as a natural consequence the definition:

\begin{equation}
\Lambda(N,x_{0}) \equiv \sum_{n=0}^{N-1} ln \bigg| \, \frac{\Delta x(n+1)}{d} \, \bigg|
\label{benettin}.
\end{equation}
For practical purposes this approach is better than the former, since
in this case it is not necessary to set $\Delta x(0)$ arbitrarily
small to ensure that at any time step the modulus of $\Delta x(n)$ is, in turn,
so small as to safely adopt the linearization assumption behind the
definition itself of the Lyapunov coefficient. Thus, in this case we
are naturally led to
\begin{equation}
\Lambda(N,x_{0}) = \sum_{n=0}^{N-1} ln \mid \Phi^{'}(x_{n}) \mid .
\label{logsum}
\end{equation}

As earlier mentioned, the proposal of Eq.(\ref{product}) rests on the use of only one auxiliary trajectory, and the proposal of Eq.(\ref{logsum}) involves the use of infinitely many auxiliary trajectories. In the ordinary extensive case here under discussion these two distinct proposals yield the same result thanks to the property: $ln(a\cdot b) = lna + ln b$. As we shall see in a moment, this equivalence is broken in the non-extensive case.

In the case of fractal dynamics\cite{TPZ97}
the proper entropic index $Q$ gets values different from
the prescriptions of ordinary statistical mechanics. This means
that $Q \neq 1$. We thus find natural to adopt
a mobile entropic index $q$, which in principle, can run
from $q = -\infty$ to $q = +\infty$. The necessity of
considering values $q \neq 1$ breaks the equivalence between
Eq. (\ref{logsum}) and Eq. (\ref{product}). Note that the entropy of Eq. (\ref{entropyt})  implies  the conventional $ ln x$ to be replaced by $ ln_{q} x \equiv \frac{x^{1-q}-1}{1-q}$. Thus, we see that the $q$-generalization of  Eq. (\ref{product}) is:

\begin{equation}
\Lambda_{q}^{(\Pi)}(N,x_{0}) \equiv
\frac{1}{1-q} \left[ \left( \prod_{n =0}^{N-1} \mid \Phi^{'}(x_{n}) \mid^{(1-q)} \right) -1 \right].
\label{qproduct}
\end {equation}
We see also that the $q$-generalization of Eq. (\ref{logsum}) is:
\begin{equation}
\Lambda_{q}^{(\Sigma)}(N,x_{0}) \equiv
\frac {1}{1-q} \left[ \sum_{n =0}^{N-1} \left( \mid \Phi^{'}(x_{n}) \mid^{(1-q)} -1) \right) \right].
\label{qsum}
\end{equation}
It is evident that the two definitions are equivalent only in the case
$q=1$.

At first sight one would be tempted to believe that the definition of
Eq.(\ref{qsum}) is more convenient that the other. In fact,
assuming ergodicity, in the limit
$N \rightarrow \infty$ this definition would make
the ratio $\Lambda_{q}^{(\Sigma)}(N,x_{0})/N$ become identical
to the average over the invariant distribution of the following form
of local Lyapunov coefficient

\begin{equation}
\lambda^{(\Sigma)}_{q} \equiv  \frac{1}{1-q} \Bigg[ \frac{\int\limits_{0}^{1} \mid \Phi^{'} (x) \mid ^{(1-q)} \rho (x) dx}{\int\limits_{0}^{1}\rho (x) dx} - 1\Bigg].
\label{wrong}
\end{equation}
However, if this definition were adopted, in the time asymptotic limit the time dependence of $ \Lambda _{q}^{(\Sigma )} (N,x_{0})$ would be linear regardless of the value of $q$. For this reason it is not convenient to use the proposal of Eq. (\ref{qsum}). In Section III we shall see that Eq. (\ref{qproduct}) is the correct proposal.

Let us get a preliminary acquaintance  with the definition
of Eq. (\ref {qproduct}). Let us
note, first of all, that 
\begin{equation}
\Lambda_{q}^{(\Pi)}(N,x_{0}) =
\frac {1 - e^{-(q-1)\Lambda(N,x_{0})}}{(q -1)}.
\label{relation}
\end{equation}
This can be easily obtained by evaluating the logarithm of the product
appearing in Eq. (\ref{qproduct}): In fact this allows us to establish
immediately a connection with the time dependent Lyapunov coefficient
defined by Eqs. (\ref{product}) and (\ref{logsum}).
 For statistical purposes it is
convenient to properly average the expression of
Eq. (\ref{relation}) over the invariant distribution.
On an intuitive basis one  would expect that the proper entropy time
evolution is given by
\begin{equation}
H_{q}^{*}(N) \:  \equiv \:  <\Lambda_{q}^{(\Pi)}(N,x)>_{q},
\label{KST}
\end{equation}
with the averaging defined by:
\begin{equation}
<\ldots>_{q} \: \equiv \: \int dx \rho(x)^{q}.
\label{properaveraging}
\end{equation}
The problems to settle are two. First of all we have
to give a stronger support to the heuristic definition
of generalized Lyapunov coefficient of Eq. (\ref{relation}).
Then we have to establish a
criterion, based on dynamics, to decide which value to assign to the mobile index $q$.
All this will be explained in Section III.

\section{generalization of the Pesin theorem}
To properly appreciate the importance of the expression of
Eq. (\ref{KST}) (supplemented by Eq. (\ref{properaveraging})) the
reader should first of all recall the importance of the
Kolmogorov-Sinai (KS) entropy\cite{K,S}. The authors of Refs.\cite{TPZ97},
\cite {costa} and \cite{lyra} have already pointed out the importance of
expressing the KS entropy in terms of the
Tsallis entropy \cite{CONSTANTINO88}, rather than
in terms of the Gibbs entropy. However, they did not go through a
derivation entirely based on a single trajectory.
The generalization of the KS entropy within the single-trajectory
spirit of the Kolmogorov approach involves the evaluation of the
following form of entropy:
\begin{equation}
H_{q}(N)\equiv \frac{1-\sum_{\omega_{0}...\omega_{N-1}}
p(\omega_{0}...\omega_{N-1})^{q}}{q-1},
\label{kolmogorov}
\end{equation}
where $p(\omega _{0}...\omega _{N-1})$ is the probability of finding the
cylinder corresponding to the sequence of symbols $\omega _{0}...\omega
_{N-1}$\cite{beck}. This sequence is generated by dynamics in the
following sense. First of all, the phase space is divided into cells, each of them is
assigned a label $\omega_{r}$, then we run a virtually
infinitely extended trajectory, with $M$ time steps, and
with $M>>N$. This means that the trajectory is much longer that the
time range explored by the subsequent entropic analysis.
At any time step this trajectory occupies a given cell, with a given
label, thereby generating a virtually infinite sequence of symbols.
This ``infinite'' sequence is explored
with a moving window of size $N$, and for any window position a
string $\omega_{0}...\omega_{N-1}$ is detected. We have to establish
the frequency of occurence of this string, to determine thus the
corresponding probability $p(\omega _{0}...\omega _{N-1}$), which is
then used to fix the value of the entropy of Eq. (\ref{kolmogorov}). This
is the reason why we say that this entropy is generated by dynamics.

The KS entropy is an entropy per unit of time. This means that the
proper generalization of the KS entropy should be given by
\begin{equation}
h_{q} \equiv lim_{N \rightarrow \infty} \frac{H_{q}(N)}{N}.
\label{realkolmogorov}
\end{equation}
We are thus in the right position to address the problem of
definining in a non-ambiguous way the earlier mentioned
\emph{dynamic entropic index} $Q$. If we chose an arbitrary $q$ the existence
of the limit of Eq. (\ref{realkolmogorov}) would not be guaranteed. In the case
where we set $q = 1$ the generalized definition of KS entropy becomes
identical to the ordinary definition, and the KS entropy is related
to the positive
Lyapunov coefficients of the trajectory by means of the Pesin
theorem\cite{pesin}. If this limit does not exist, namely, it is either
a vanishing or an infinite quantity, one would be tempted to conclude
that the dynamic process under study is incompatible with
thermodynamics in the sense of Kolmogorov. However, if it happens
that a given value of $q$ exists,
such that the limit of Eq. (\ref{realkolmogorov})  exists and is finite, then
we consider it to be the dynamic entropic index of the process under
study. We call $Q$ this ``magic'' value of $q$, which is in fact the
proper dynamic entropic index of the process.
In an earlier work\cite{luigi}
a numerical calculation was made to establish $Q$ in the case of a text
of only two symbols, with strong correlation.
The case of many more symbols is beyond the range
of the current generation of computers.

 This arduous task is made less difficult by the fact that recently
 a proper generalization of the Pesin theorem has been
 established\cite{jin}. According to the prescription of this paper
 the entropy
 $H_{q}(t)$ of
Eq.(\ref{kolmogorov}) reads

\begin{equation}
H_{q}(t)\equiv \frac{1-\delta ^{q-1}\int dxp(x)^{q}\xi (t,x)^{1-q}}{q-1},
\label{jin}
\end{equation}
where the symbol $t$ denotes time regarded as a continuous variable.
In fact, when
the condition $N>>1$ applies, it is legitimate to identify $N$ with $t$. The
function $p(x)$ denotes the equilibrium distribution density and $\delta $
the size of the partition cells: According to \cite{jin} the phase
space, a one-dimensional interval, has been divided into $ W = 1/\delta $
cells of
equal size.

We have to remark that the reason why the Kolmogorov entropy is so
popular is because it affords a way of establishing the randomness of
a process without any arbitrary dependence on the way the observation
process is carried out. In fact, in the case where the partition of
the phase space into cell is generating \cite{beck}, the resulting
entropy rate is independent of the partition adopted. This important
aspect is retained by the generalized expression of Eq.(\ref{jin}). In
fact, this equation apparently suggests  that the independence of the
entropy rate of the partition adopted is guaranteed only when $q =
Q = 1$. Actually, a recent work\cite{simone} has assessed that if the
invariant distribution is multifractal the expression of Eq. (\ref{jin})
becomes independent of the size of the cells, and the proper dynamic
entropic index can be different from unity. On the other hand, this means
that in the
case where the invariant distribution is smooth, rather than
multifractal, the dynamic entropic index $Q$ must become equal to the
unity. We shall discuss the physical consequences of this important
conclusion in the next sections.

 \section{the Manneville map}

 The
  Manneville map \cite{manneville}
 reads:
  \begin{equation}
  x_{n+1} = \Phi(x_{n}) = x_{n} + x_{n}^{z} (mod 1) \quad (z \geq  1).
  \label{mannevillemap}
  \end{equation}
It is characterized by a laminar region given by the
  interval $[0,d]$ with $d$ defined by the equation:
  \begin{equation}
  d^{z}+d = 1  .
  \label{ddefinition}
  \end{equation}
  It was recently shown\cite{ANNA} that in this
  case the prediction of Eq. (\ref{unconventional}) holds true with
  \begin{equation}
  Q = 1 + (z-1)/z .
  \label{magicQ}
  \end{equation}
 
 Note that the condition $ z = 1$, making the Manneville map equivalent to the Bernoulli shift map [1], yields, as it must, $Q = 1 $ and thus a form of ordinary statistical mechanics. It is easy, as well as convenient, to write explicitly the two Lyapunov coefficients of Eqs. (\ref{qproduct}) and (\ref{qsum}) in this case. Due to the fact that $ \frac{d\Phi (x)}{dx} = 2 $ we obtain that the two prescriptions result in

\begin{equation}
\label{qproductshift}
\Lambda_{q}^{(\Pi)} (N) = \frac{1 - 2^{N(1-q)}}{q - 1}
\end{equation}
and

\begin{equation}
\label{qsumshift}
\Lambda_{q}^{(\Sigma)} (N) =  N \:  \frac{2^{(1-q)} - 1}{(1-q)},
\end{equation}
respectively. We see that Eq. (\ref{qproductshift}) results in a linear time evolution of the Lyapunov coefficent  only when $ q = Q = 1 $. For $ q > Q $  the Lyapunov coefficent  increase is slower and for $ q < Q $ is faster. A steady entropy increase per unit of time can only be defined for $ q = Q = 1$. We make therefore the conjecture that this property migth hold true even when $ Q \neq 1 $, in agreement with the result of numerical work of  \cite{luigi}. We also see, as already pointed out in Section II, that Eq. (\ref{qsum}) yields Eq. (\ref{qsumshift}) and this time dependent Lyapunov coefficent turns out to be linear in $N$ regardless of the value of $q$ used.

 To help the reader to understand the significance of the results of
  this paper, it is convenient to review the physical arguments used
  in \cite{ANNA} to make the
  prediction of Eq. (\ref{magicQ}).
  The authors of \cite{ANNA} used an intermittent map with two
  laminar regions, each of which is equivalent to the laminar region of
  the Manneville map. The sojourn in one of the two laminar regions
  means uniform motion in the laboratory frame of reference with
  velociy $W$ in either the positive or the negative direction. The
  random walker created by this dynamic process makes jumps of
  intensity $|x|$ in either the positive or the negative direction
  with a probability $\Pi(|x|)$, which is related to the waiting time
  distribution $\psi(t)$ in one of the two equivalent laminar regions
  by means of the key relation
  \begin{equation}
  \Pi(|x|) = \frac {\psi(\frac{|x|}{W})}{W}.
  \label{jumping}
  \end{equation}
  On the other hand the intermittent map used in \cite{ANNA}
  yields a waiting time distribution $\psi(t)$ whose explicit
  expression is:
  \begin{equation}
  \psi(t) = d^{z-1} [1 + d^{z-1}(z-1) t]^{\frac{z}{1-z}}.
  \label{waiting}
  \end{equation}

We will focus our attention on the interval $ 1.5 < z < 2$ of the map parameter.
This is the range of $z$ for which only the first moment  of $\psi(t)$ is finite and anomalous diffusion is expected \cite{ANNA}. For $ 1 < z < 1.5 $ also the second moment of $\psi(t)$ is finite and ordinary diffusion is obtained. Finally, in the case of $z>2$ all the moments are infinite, thereby conflicting with the requirement of producing diffusion compatible with an Hamiltonian description \cite{zasla}.

 The maximization of the non-extensive entropy $S_{q}(\Pi(|x|))$,
 namely, the non-extensive entropy  expressed as a functional
 of $\Pi(|x|)$ yields an analytical
 expression for $\Pi(|x|)$ which is compatible with the inverse
 power law form inherited by $\Pi(|x|)$ through Eq. (\ref{jumping}).
 This is the origin of the prediction of Eq. (\ref{magicQ}).
 The authors of Ref. \cite{ANNA}, using dynamic arguments, prove that the
Manneville map yields
 \begin{equation}
 \xi(t) = [1 - (z-1)x^{z-1}t]^{-\frac{z}{z-1}}.
 \label{crucialdifficulty}
 \end{equation}
 As far as the power index is concerned, this dynamical property is
 compatible with Eq. (\ref{unconventional}) and with the prediction
 of Eq. (\ref{magicQ}). Unfortunately, the coefficient $\lambda_{Q}$
 of Eq. (\ref{unconventional}) through comparison with
 Eq. (\ref{crucialdifficulty}) turns out to be
 \begin{equation}
 \lambda_{Q} = z \, x^{z-1}.
 \label{baddependence}
 \end{equation}
 This dependence on the initial condition conflicts with the fact
 that the waiting time distribution is independent of the initial
 condition, being a  statistical average based on the assumption of a
 random but uniform injection from the chaotic to the laminar
 region\cite{zumofen}.

 The rationale for this conflict is that the generalized Pesin
 theorem of Section III implies that the dynamic entropic index gets
 the ordinary value $Q=1$. In fact, it is well known\cite{gaspard}
 that the invariant distribution is given by
 \begin{equation}
 \rho(x)  = \frac{const} {x^{z-1}}.
  \label{invariant}
  \end{equation}
  This means a smooth distribution, and consequently according to the
  arguments of Section III, $Q = 1$. This prediction seems to be in a
  harsh conflict with the prediction of Eq. (\ref{magicQ}). However, as
  we shall see in the next section, it is not so, since the prediction
  of Eq. (\ref{magicQ}) refers to a microscopic process with
  high memory, before the crossing from the laminar to the chaotic
  region occurrs.

 \section{numerical results}
  The prediction $Q=1$ of the generalized version of the Pesin theorem
  of Eq. (\ref{jin}) refers to a condition of equilibrium with the
  system in its invariant distribution. To establish a contact with
  the prediction of Eq. (\ref{magicQ}) it is convenient to average
  the Lyapunov coefficient of Eq. (\ref{qproduct}) on a non-equilibrium
  distribution density. Thus we use the numerical work to evaluate
  \begin{equation}
  A_{q}(N) \equiv <\Lambda_{q}^{(\Pi)}(N,x)>_{ne}.
  \label{nonequilibrium}
  \end{equation}

Note that this prescription has not to be confused with that of Eq. (\ref{KST}) even if both are prescriptions to make an average over the same Lyapunov coefficient, that of Eq. (\ref{relation}). Here, $ <...>_{ne} $ denotes an average different from that of Eq. (\ref{properaveraging}), since we use $ \rho (x)$ rather than $ \rho^{q} (x) $. Furthermore the statistical weight $ \rho (x)$ is not the natural invariant distribution of Eq. (\ref{properaveraging}). We select on the contrary, as a statistical weight, a uniform distribution from $ x = 0$ to $ x = \Delta  < d $. All the numerical calculations of this section refer to $ \Delta = 10^{-4}$. Note that, as discussed in  \cite{ANNA}, the range of $z$-values corresponding to the emergence of L\'{e}vy processes is given by the interval $ \left[1.5,2 \right]$. We select the intermediate value $ z = 1.7 $ and, for computational convenience, the Lyapunov coefficent of Eq. (\ref{relation}) is expressed in the equivalent form

  \begin{equation}
  \Lambda_{q}^{(\Pi)}(N,x) = \frac{1 - \xi(N,x)^{(1-q)}}{q-1}.
  \label{computationalpurposes}
  \end{equation}
We aim at establishing at any given time region the value of
the entropic index $q$ which makes the corresponding non-extensive
entropy increase linearly  with N. This is a temporary "magic" Q, since
, as we shall see, the windows of linear increase have a finite time
duration. We study the time derivative of $A_{q}(N)$ of Eq. (\ref{nonequilibrium}), namely:
  \begin{equation}
  B_{q}(N) = A_{q}(N+1) - A_{q}(N).
  \label{derivative}
  \end{equation}
The purpose of the numerical calculation is that of determining the time regions where $ B_{q} (N)$ is constant.

The theoretical remarks of Section IV imply that only one proper "thermodynamical" region exists, this being given by $ B_{q} (N)$ constant for an infinite time interval. This is the region of ordinary statistical mechanics with $ Q = 1$. However, we shall show that the time evolution of the Manneville map ensuing a given off-equilibrium initial condition goes through non-extensive phases of increasingly time duration, and this time duration becomes larger and larger as the deviations from the extensive condition become slighter and slighter.

In Fig. 1 we plot the quantity $ B_{1} (N) $. Note that $ B_{1} (N) $ is the difference between two subsequent iteration steps of the Lyapunov coefficient $\Lambda(N,x)$ of
Eq. (\ref{product}) averaged over a non equilibrium distribution. We see that a relatively extended time region shows up, ranging from $ N = 0$ to $ N \simeq 900 $, where $B_{1}(N)$ is approximately constant with an almost vanishing value. This is the region where according to the rule established intuitively with the help of Eq. (\ref{qproductshift}) the existence of a vanishing non-extensive Lyapunov coefficient at a given $q$ implies that a finite non-extensive Lyapunov coefficient might exists at a larger value of $q$. In fact, as we shall see with the help of Fig. 3, at the magic value of $q$ given by Eq. (\ref{magicQ}), this time region is found to correspond to a finite Lyapunov coefficient.

After this initial time region $B_{1}(N)$ undergoes an abrupt
increase followed by a slow regression to a constant non vanishing value. This
fits very well the theoretical prediction of Section IV. The slow regression corresponds to the time evolution of the initial non-equilibrium distribution towards the final invariant distribution. This has the effect of realizing a condition equivalent to that of Eq. (\ref{pesin}). In conclusion, Fig. 1 sheds light into the process of transition to equilibrium , and gives a further evidence to the fact that this equilibrium is of extensive nature. This figure suggests that the final equilibrium condition is reached at about $ N \simeq 15,000$. However, as we shall see with the help of Figs. 2, 4, 5 and 6, at this time the process of regression to equilibrium is still active.


\begin{figure}
\begin{center}
\epsfig{file=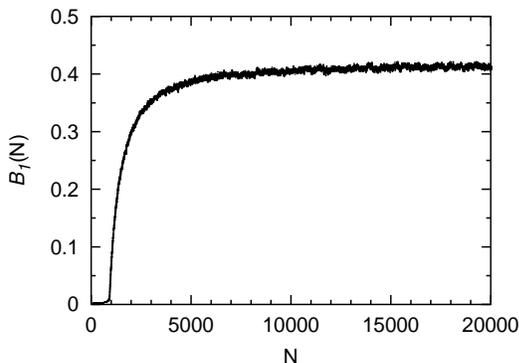,height=7cm,width=5cm,angle=270}
\caption{{\it B}$_{1} (N)$ as a function of the iteration step $N$.  $ {\it B}_{1} (N)$ has a vanishing value until $ N \simeq 900$ after which  undergoes an abrupt increase followed by a slow regression, with statistical fluctuations, to the constant value predicted by Pesin theorem.}
\end{center}
\end{figure} 

We think that a condition of genuine regression to equilibrium implies that all the trajectories, departing from the selected off-equilibrium initial distribution, have crossed the border between the laminar and chaotic region at least once. The process of regression to equilibrium is slow because the motion of trajectories close to $ x = 0 $ is slow. To shed light on this important aspect, let us adopt the continuous time approximation to Eq. (\ref{mannevillemap}) and let us write:

\begin{equation}
dx/dt = x^{z}.
\label{continuoustime}
\end{equation}
In line with the prescription of denoting by $t$ the discrete time $N$ when
$N>>1$ applies, from here throughout the remainder of this Section, we shall be
denoting time with symbol $t$. The solution of Eq.(34) for a trajectory with
initial condition x(0) is given by

\begin{equation}
x(t) = [x(0)^{1-z} - (z-1)t]^{\frac{1}{1-z}}.
\label{solution}
\end{equation}

  Using this solution it is easy to find the time at which the first
  trajectory, that belonging to the right border of the initial distribution,
  exits from the laminar region. This time, denoted by $T$, is given by
   \begin{equation}
  T = \frac{d^{1-z}}{z-1} [(\frac{\Delta}{d})^{1 - z} - 1].
  \label{time}
  \end{equation}
  Before this time no trajectory can exit from the laminar region.
  To evaluate the population decrease after this time,
  we notice that:
     \begin{equation}
  dM = M(0) \frac{dx}{\Delta},
  \label{starting}
  \end{equation}
  where $M(0)$ is the number of trajectories within the laminar region at
  the initial time. Thus, using Eq. (\ref{solution}) we obtain
  \begin{equation}
  \frac{dM}{dt} = -\frac{d^{z}}{\Delta} \frac{M(0)}{[1 + 
   d^{z-1} (z-1) t]^{\frac{z}{z-1}}},
  \label{equationofmotion}
  \end{equation}
  which makes it possible for us to establish the time evolution of
  the population at time $t$ for $t>T$. In conclusion, we get
   \begin{equation}
  M(t) = M(0)   \quad   (t < T)
  \label{before}
  \end{equation}
  and
    \begin{equation}
  M(t) = M(0)  \frac{d}{\Delta} \frac {1} {[1 +
  d^{z-1}(z-1)t]^{\frac{1}{z-1}}} \quad (t > T).
  \label{after}
  \end{equation}
It is interesting to remark that the time derivative of $M(t)$, at $t > T$, as
resulting from Eq. (\ref{after}), turns out to be proportional to the waiting time
distribution $\psi(t)$ of Eq. (\ref{waiting}). This means that $M(t)$, although depending on
an arbitrary initial condition, at times larger than $T$ reflects the stationary
and statistical nature of $\psi(t)$. This fits the numerical observation made
herein with the help of Fig. 6 that the process of regression to equilibrium is
not affected by the return of the trajectories from the chaotic to the laminar
region. We shall see, in fact, that the long-time behavior of $M(t)$ of Eq. (\ref{after}) is
a fair indicator of the process of regression to equilibrium. 

In Fig. 2 we plot $M(t)/M(0)$ as resulting from the numerical treatment. The result obtained coincides with the theoretical prediction of Eqs. (\ref{before}) and (\ref{after}). It looks like the
mirror image of the curve of Fig. 1, thereby confirming that the
relaxation of the non-equilibrium  Lyapunov coefficient is due to
the trajectories crossing at least once the border between laminar
and chaotic region.


\begin{figure}
\begin{center}
\epsfig{file=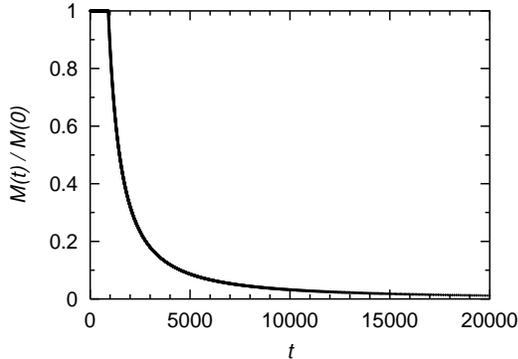,height=7cm,width=5cm,angle=270}
\caption{Time evolution of the population of the laminar region. The numerical result coincides with the time evolution predicted by Eqs. (39) and (40) and the initial plateau, lasting for the time  $ T \simeq 900$ coincides with the prediction of  Eq. (36).}
\end{center}
\end{figure}


\begin{figure}
\begin{center}
\epsfig{file=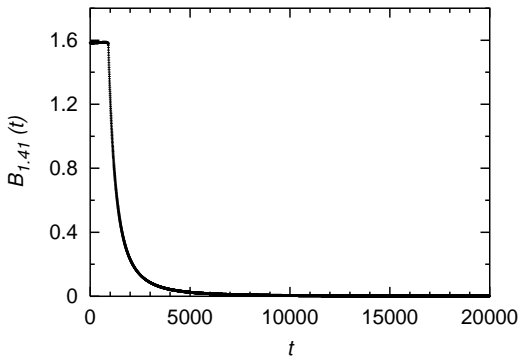,height=7cm,width=5cm,angle=270}
\caption{$ B_{1.41}$, multipied by the factor of $10^{3}$, as a function of time. The value  of Q predicted by Eq. (\ref{magicQ}) turns out to be the proper entropic index in the initial time interval $ \left[0,T \right]$ with  $ T \simeq 900$. The subsequent quick drop to zero signals the regression to $ q = 1 $. }
\end{center}
\end{figure}
 
In Fig. 3 we study the time derivative of the Lyapunov coefficient
  of Eq. (\ref{qproduct}) with $q = Q = 1.41$, that is, the value obtained from Eq. (\ref{magicQ}) with $ z = 1.7$. We see that this rate of entropy increase per time unit is constant in the initial interval $ 0 < t <  T $. This means that the non-extensive entropy corresponding to the magic value of the entropic  index $q$ given by the theoretical prediction of  \cite{ANNA}, Eq. (\ref{magicQ}), realizes, as it must do, the Kolmogorov condition of an increase linear in time. From Fig. 3 we also see that after the first escape from the laminar region, at $ t = T$, the rate per unit time of this non-extensive entropy drops quickly to much smaller values with a subsequent slower relaxation to a vanishing value in the time asymptotic limit. This means that in this time region the time increase of this non-extensive entropy becomes slower than a linear function in time, thereby suggesting, on the basis of the intuitive rule established by Eq. (\ref{qproductshift}), that the proper entropic index in this case is smaller than that given by the prediction of Eq. (\ref{magicQ}). In fact, we have already argued that at equilibrium $ Q = 1$. 


\begin{figure}
\begin{center}
\epsfig{file=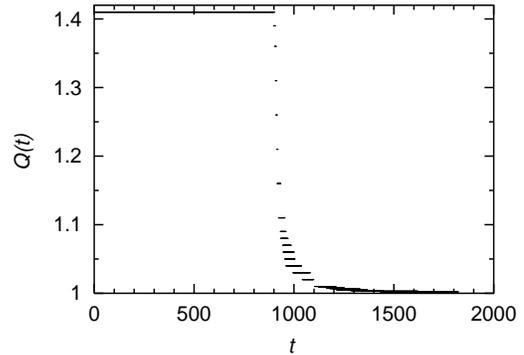,height=7cm,width=5cm,angle=270}
\caption{ Q as a function of time: The fast drop at $ t = T $. We see that at $ t = T $, when the first trajectory escapes from the laminar region, $ Q(t) $ drops from the value $1.41$, established by Eq. (\ref{magicQ}), to the value $1.01$ in about $200$ iteration steps. This fast drop is followed by a much slower relaxation to the final value of $1$, detailed in Fig. 5.}
\end{center}
\end{figure}

In Figs. 4 and 5 we show the change of the entropic index from the prediction of
Eq.(\ref{magicQ}), $Q = 1 + (z-1)/z$ ,  corresponding to a trajectory motion not
yet affected by the sporadic randomness, to the asymptotic value $Q=
1$, produced by the repeated action of the chaotic region of the phase
space. In section III we have seen that it is the smooth nature of the
invariant distribution that makes it impossible to adopt the general
prescription of Eq. (\ref{jin}) in a form different from that of the ordinary
Pesin theorem. On the other hand, this invariant distribution is
realized through the process of escape from the laminar region, and so
by the action of sporadic randomness, whose effect is illustrated here
in detail. The calculation is done by searching, at any time $t > 0$,
for the value of $q$ realizing, temporarily, the condition of linear
increase of the non-extensive entropy. This, in turn, is realized by
looking for the value of $q$ making $B_{q}(t)$ of Eq. (\ref{derivative}) constant over
windows of finite size. The window sizes are reported in both figures
as small intervals, whose length tends to increase as the value of $Q$
decreases. Fig. 4 shows the remarkable fact that the transition from
the region where only order exists to that where sporadic randomness
begins showing up is signalled by the fast drop of the entropic index
$Q$.

Nevertheless after the fast drop at $ t = T $, the equilibrium value $ Q = 1$ is reached asymptotically in time with a slow regression process. This is illustrated by Fig. 5, which shows more clearly
than Fig. 4, that the time duration of this temporary non-extensive thermodynamics becomes larger and larger as $Q$ comes closer and closer to the equilibrium value $ Q = 1$. We think that the adoption of the non-extensive formalism results in an impressively sensitive indicator of the slight departure from the final equilibrium distribution. 


\begin{figure}
\begin{center}
\epsfig{file=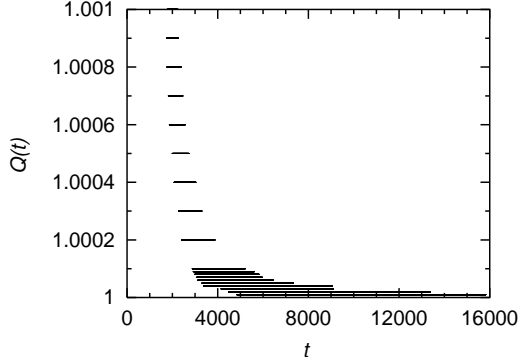,height=7cm,width=5cm,angle=270}
\caption{Q as a function of time: Details of the slow regression following the fast drop. This figure shows, more clearly than Fig. 4, the windows of entropy linear increase. The size of these windows becomes larger and larger at greater and greater times. The largest window shown here corresponds to $Q = 1.00001$. In the scale of this figure smaller values of $q$ becomes indistinguishable from the value $ q =1 $, which corresponds to a window of infinite time size.}
\end{center}
\end{figure}

The process of regression of $Q(t)$ to the equilibrium value $Q=1$ is closely related to the process of transition from the initial unstable distribution to the final invariant measure. This latter process, in turn, is determined by the trajectories exiting the laminar region, and consequently is related to the time evolution $ M(t)$ of Eqs. (\ref{before}) and (\ref{after}). However, an essential part of this process of relaxation to equilibrium might be played also by the trajectories that from the chaotic region are injected back into the laminar region. To establish if this true or not, it is convenient to
monitor numerically the process of transition to equilibrium. This is done as follows. The interval $ \left[0,1 \right]$ is divided into $C$ cells of equal size. In our calculations we set $C=100$. We consider $M$ trajectories with initial conditions uniformly distributed over the whole interval $ \left[0,1 \right]$ and we iterate all of them N times. We set $M= 10,000 $ and $N = 100,000 $. In accordance with the prescription of Section III, we denote time with the symbol $t$. At this stage we evaluate how many trajectories are found in a given cell with the label $i$. We call $M_{i}$ this number and we set  $P_{i}^{eq} =  M_{i}/M$. This stage does not afford yet a proper numerical determination of the equilibrium distribution. This is so because with a finite number of trajectories $M$  and cells of a finite size $1/C$, the quantity $P_{i}^{eq}$ will turn out to be a fluctuating function of the iteration time $t \equiv N$. The intensity of these fluctuations depends on the selected values for the numbers $C$ and $M$. To bypass this limitation we make a time average on $\tau$ further iterations after the time $t$, thereby defining
\begin{equation}
\label{peqtau}
< P_{i}^{eq}>  _{\tau} = \frac{1}{\tau} \sum _{t^{'}= \; 0}^{\tau} P_{i} (t^{'}) ;
\end{equation}
where  $< P_{i}(t)>  _{\tau}$ denotes the probability of the $i$-th cell  $t^{'}$  further iterations after the time $t$. The result of this calculation is scarcely
dependent on the value of tau adopted if this is much greater than the
time scale of the fluctuations of $P_{i}^{eq}$: In our case for $\tau > 500$ the
right hand side of Eq. (\ref{peqtau}) remains practically constant for all the cells. Thus, we can omit the dependence of $< P_{i}^{eq}>  _{\tau}$ on $\tau$ and use Eq. (\ref{peqtau}), with a given value of $\tau > 500 $, to define our numerical equilibrium distribution, which, for the sake of simplicity is again denoted  by the symbol $ P_{i}^{eq} $.

Now we can address the important issue of the regression to equilibrium. First of all, we adopt the same initial condition as that used in the earlier calculations. From this initial distribution we select a sample of $M$ trajectories. At any time step we count how many trajectories are found in a given cell, thereby determining $P_{i}(t)$, namely, the probability that a trajectory is found in the cell with label $i$ at time $t$. We compare this probability with the equilibrium probability, evaluated according to the earlier numerical prescription, thereby defining the variable $Y_{i}(t)$ as

\begin{equation}
\label{difference}
Y_{i} (t) = \; \mid P_{i} (t) -  P_{i}^{eq} \mid \; .
 \end{equation}
Then we evaluate the relative dispersion $R_{i} (t)$ of the quantities $ Y_{i} (t)$ around the equilibrium value $ P_{i}^{eq} $

\begin{equation}
\label{revdisp}
R_{i} (t) = \frac{Y_{i} (t)}{P_{i}^{eq}} .
\end{equation}

We have now to deal with the issue of the fluctuations of $R_{i} (t)$ caused by the adoption of finite values for $C$ and  $M$. We follow the same procedure as that adopted to determine the equilibrium distribution. This means that we make the time average 

\begin{equation}
\label{avgtimedifference}
< Y_{i}(t) > _{\tau} = \frac{1}{\tau} \sum _{t^{'}= \; t}^{t\; +\; \tau} Y_{i} (t^{'}) ,
\end{equation}
thereby deriving the time average of the relative dispersion $R_{i} (t)$:

\begin{equation}\label{avgreldsp}
< R_{i} (t) > _{\tau} \, = \, \frac{< Y_{i} (t) >_{\tau}}{P_{i}^{eq}} .
\end{equation}

We, then, consider a set of numbers $R$ in the interval $ [0,1] $.For each
of these numbers we determine the number of iterations $T$ necessary to
make the time average of the relative dispersion,  Eq.  (\ref{avgreldsp}),  smaller
than $R$,for all the $C$ cells. This makes it possible for us to use the function $R(T)$ as a fair indicator of the relaxation to equilibrium. In fact, for any time $T$ we can say that the corresponding distribution departs from equilibrium by an amount of the order of $R*100$ per cent. The values of $\tau$ adopted range from $\tau = 500$ to $\tau = 3000$. Within this wide interval the change of $R(T)$ is not significant.

In Fig. 6 we plot R(T) for different values of
$\tau$. This figure shows that at $T=10,000$ and $T=12,000$ the distribution stills  departs from equilibrium, for all the values of $\tau$, by a quantity of the order of $3\sim 4$ per cent.We need to wait till to a time greater than $20,000$ to detect a smaller departure from equilibrium. Consequently, according to the criterion we have adopted, at $T=16,000$ the departure from equilibrium is expected to be still of the order of $3\sim 4$ per cent.

It would seem to be plausible to make the conjecture that the return of the trajectories from the chaotic to the laminar region makes the relaxation to equilibrium  slower than the decay of  $M(t)$ given by Eq. (\ref{after}). This is so because we arrived at this expression for $M(t)$ by neglecting the process of return of the trajectories to the laminar region after the escape into the chaotic region. However, the numerical calculation reported in Fig. 6 shows that it is not so and that $M(t)$ is a good indicator of the process of
relaxation to equilibrium. Thus, we can use $M(t)$ to establish if the
small departures of $Q(t)$ from $Q = 1$ are a genuine indication of memory
of the initial condition. For example the long window of linearity
between $t \approx 6,000$ and $t \approx 16,000$, corresponding to $Q = 1.00001$ , is shown
by  $M(t)$ to correspond to a departure from equilibrium of the order of $3\sim 4$ per cent. This means that this very small departure from the
ordinary condition $Q = 1$ signals that about $3\sim 4$ per cent of the
trajectories are still in the laminar region.


\begin{figure}
\begin{center}
\epsfig{file=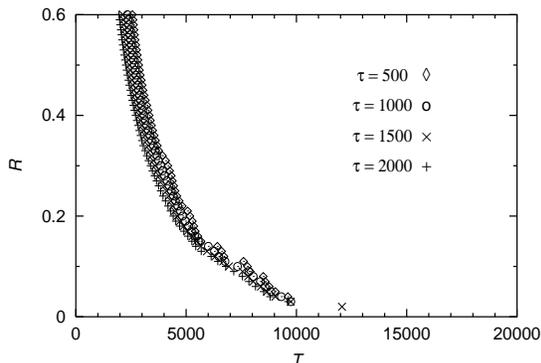,height=7cm,width=5cm,angle=270}
\caption{$R$ as a function of the time $T$. Here the number of cells C is 100  and the trajectories used ,$M$, are 10000. }
\end{center}
\end{figure}

\section{concluding remarks}
This paper illustrates the dynamical process of the memory erasure necessary to realize the processes of L\'{e}vy diffusion, and therefore makes it possible to complete the program of the work of \cite{ANNA}. In fact, the authors of that paper focused their attention on the microscopic process responsible for anomalous diffusion of L\'{e}vy kind. This implies elementary jumps of a given length $x$ with a probability proportional to $1/|x|^{\frac{z}{z-1}}$. This requires the maximization of the non-extensive entropy\cite{ANNA} with the entropic index given by Eq.(\ref{magicQ}). However, when an attempt is made at establishing a connection between this extensive property and the KS entropy, either in the normal or in the non-ordinary form\cite{jin}, a conflict emerges due to the fact that according to the work of \cite{gaspard} the ordinary version of the Pesin theorem applies.

Actually the process of regression to equilibrium is shown to be very slow and take place according to the theoretical prediction of Eqs. (\ref{before}) and (\ref{after}). We do not find any numerical evidence supporting the plausible conjecture that the process of regression to equilibrium is made slower by the process re-injection of the trajectories into the laminar region. In conclusion, as an important result of this paper, we show that there is a close connection between this slow process of regression and the aging of the entropic index: The process of aging corresponds to the memory erasure necessary to establish the L\'{e}vy statistics \cite{mauro}.

This process of memory erasure, beggining at $ t = T \simeq 900$, is not as fast as it seems to be at a first sight. The numerical and theoreticaal arguments of Section V prove that it is slow. In other words, a very extended regime exists, of balance between the randomness of the chaotic region and the order of the laminar region, and it is signaled by values of the entropic index $Q$ very close to $ Q = 1$, but not coinciding with this value. The reader might find surprising that $ Q = 1.00001 $ signals the existence of memory, in spite of being so close to the prediction of ordinary statistical mechanichs, $ Q = 1$. Actually, our theoretical and numerical arguments prove that this
very weak deviation from $Q = 1$ corresponds to a process of relaxation
still active. This conclusion, of course, is made possible by the fact that the dynamic process under study is understood since relatively long time \cite{gaspard}.

We see essentially two main benefits emerging from the results of this paper. The first is that the non-extensive entropy might serve the purpose of detecting a residual order in the long-time limit of real time series, where we would expect a totally random behavior. The second is of conceptual nature and has to do with the meaning itself of the non-extensive thermodynamics. The present paper suggests that the non-extensive thermodynamic condition is not permanent, and rather refers to non-equilibrium process of long time duration.


\begin{references}
\bibitem{hilborn} R.C. Hilborn, {\em Chaos and Nonlinear Dynamics},
Oxford University Press, New York (1994).

\bibitem{brazil} C.Tsallis, Braz. J. Phys. {\bf 29},1 (1999)\newline
[http://www.sbf.if.usp.br/WWW$_{-}$pages/\newline
Journals/BJP/Vol29/Num1/index.htm].

\bibitem{TPZ97}  C. Tsallis, A.R. Plastino, and W.-M.Zheng, Chaos, Soliton,
Fractals, {\bf 8}, 885 (1997).

\bibitem{costa} U.M.S. Costa, M.L. Lyra, A.R. Plastino and C. Tsallis,
Phys. Rev. {\bf 56}, 245 (1997).

\bibitem{lyra} M.L. Lyra and C. Tsallis, Phys. Rev. Lett. {\bf 80},
53 (1998).

\bibitem{CONSTANTINO88}  C. Tsallis, J. Stat. Phys. {\bf 52}, 479 (1988).

\bibitem{ANNA} M. Buiatti, P. Grigolini, A. Montagnini,
Phys. Rev. Lett {\bf 82}, 3383 (1999).

\bibitem{mauro} M. Bologna, P. Grigolini, J. Riccardi, Phys. Rev. E {\bf82} , 6432 (1999).

\bibitem{pesin} Ya. B. Pesin, Russian Mathematical Surveys
{\bf(4)32},55(1977).

\bibitem{jin} J. Yang, P. Grigolini, Phys. Lett. A 263, 323 (1999).

\bibitem{manneville} P. Manneville, J. Phys.(Paris) {\bf 41}, 1235
(1980).

\bibitem{schuster} H.G. Schuster,
{\em Deterministic Chaos}, Second Revised Edition,
VCH, New York (1988).

\bibitem{gaspard} P. Gaspard, X.J. Wang, Proc. Natl. Acad. Sci.
USA, {\bf 85},
4591 (1988).

\bibitem{benettin} G. Benettin, L. Galgani, Phys. Rev. A {\bf 14},
2338 (1976).




\bibitem{K} A. N. Kolmogorov, Dok. Acad. Nauk.
SSSR {\bf 119}, 861 (1958).

\bibitem{S} Ya. G. Sinai, Dok. Acad .Sci. USSR {\bf 124(4)},768 (1959).


\bibitem{luigi}
 M. Buiatti, P. Grigolini, L. Palatella,
Physica A {\bf 268}, 214 (1999) .

\bibitem{beck}  C. Beck, F. Schl\"{o}gl, {\em Thermodynamics of chaotic
systems}, Cambridge University Press, Cambridge (1993).

\bibitem{simone} S. Montangero, L. Fronzoni, P. Grigolini, cond-mat/9911412

\bibitem{zasla} G. M. Zaslavsky,  {\em Physics of Chaos in Hamiltonian Systems: on the Foundations of Statistical Physics}, Imperial College Press, London (1999)

\bibitem{zumofen} G. Zumofen and J. Klafter, Phys. Rev. E {\bf 47},
851 (1993).







\end{references}
\end{document}